\documentclass[a4paper,11pt]{article}
\pdfoutput=1 % if your are submitting a pdflatex (i.e. if you have
\usepackage{jinstpub} % for details on the use of the package, please
\title{\boldmath The Optosystem: validation and testing of the high-speed \color{black}electro-optical \color{black} conversion system for the readout of the ATLAS ITk Pixel upgrade}

 \author{S. M\"{o}bius,}
 \affiliation{University of Bern, , Laboratory for High Energy Physics (LHEP),
Albert Einstein Center for Fundamental Physics\\Sidlerstrasse 5, 3012 Bern, Switzerland}

\emailAdd{silke.moebius@unibe.ch}

\abstract{After Run III the ATLAS detector will undergo a series of upgrades to cope with the harsher radiation environment and increased number of proton interactions in the High Luminosity-LHC. One of the key projects in this suite of upgrades is the ATLAS Inner Tracker (ITk). The pixel detector of the ITk must be read out accurately and with extremely high rate. The Optosystem performs \color{black}electro-optical \color{black} conversion of signals from the pixel modules. This paper presents recent results related to the performance of the data transmission chain \color{black}with a focus on \color{black} the Optoboards and to the design, testing and production of the Optopanels.
}

\keywords{Radiation-hard detectors, electro-optical conversion, Optoboard, Optosystem, ITk, ATLAS upgrade}

\arxivnumber{} % only if you have one

\collaboration[c]{on behalf of the ATLAS ITk Group%\footnote{Copyright 2021 CERN for the benefit of the ATLAS Collaboration. Reproduction of this article or parts of it is allowed as specified in the CC-BY-4.0 license.}
}

\proceeding{Topical Workshop on Electronics for Particle Physics 2023\\
  1 October 2023 to 6 October 2023\\
  Geremeas Sardinia}

\notoc
\begin{document}
\maketitle
\flushbottom

\section{Introduction}
\label{sec:intro}
The LHC\cite{Evans:2008zzb}, a $pp$ collider at the energy frontier of 13.6\,TeV, operates at a 40\,MHz bunch crossing rate. 
It is planned to be upgraded in terms of luminosity and energy to enhance the physics reach.
The integrated luminosity of the LHC will be increased by a factor of ten and the instantaneous luminosity by a factor of 5-7.5 compared to the design luminosity of the LHC between 2028 and 2036. 
This High Luminosity-LHC (HL-LHC) has an increased particle density in the experiments, which results in a higher occupancy in the detectors.  

This poses significant challenges to the current detectors and several subsystems are replaced by radiation harder, faster detectors of higher granularity. 
The upgrade of the Inner Detector (ID) of the ATLAS experiment \cite{PERF-2007-01} to an all-silicon system, the Inner Tracker (ITk) \cite{CERN-LHCC-2017-021,CERN-LHCC-2017-005}, is the focus of this document. 
With the higher luminosity, a pile up of up to $\sim 200$ vertices in one single bunch crossing is possible.
Resulting from the increase of interaction vertices, the track density is also increased, which requires a higher granularity of the detector to keep the occupancy below 1\,\%.
Compared to the ID, the number of electrical channels increases from $\sim 10^8$ to $5\times 10^9$ in the ITk \cite{CERN-LHCC-2017-021}.
The fluence received during the detector lifetime will also be higher with $1.3\times 10^{16} \text{n}_{\text{eq}}\text{cm}^{-2}$ compared to $5\times 10^{15}\,\text{n}_{\text{eq}}\text{cm}^{-2}$ in the LHC and a foreseen maximum integrated dose of $\sim 10$\,MGy.

\begin{figure}[h!]
\centering
\includegraphics[width=.7\textwidth]{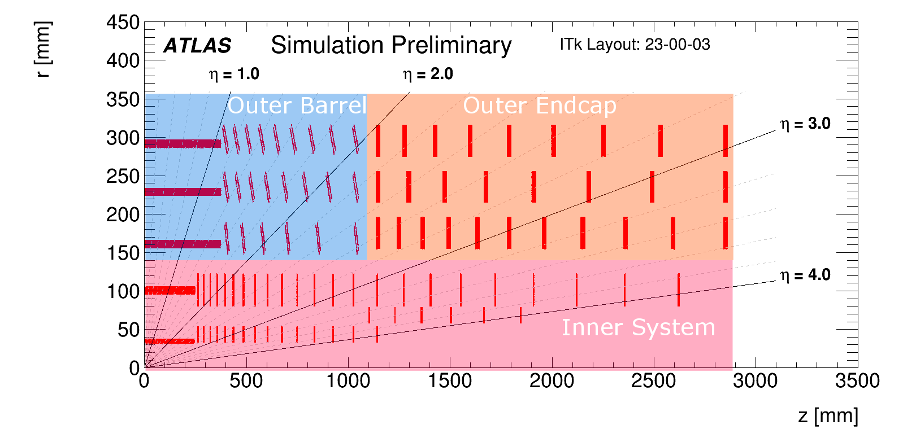}
\caption{\label{fig:itklayout} The layout of the ITk pixel detector can be divided in three subsystems, the Inner System, the Outer Barrel and the Outer Endcap (taken from \cite{CERN-LHCC-2017-005}).}
\end{figure}

The ITk features a five-layered hybrid pixel detector with planar sensors covering an area of 13\,m$^2$ and a four-layered strip detector.
Figure\,\ref{fig:itklayout} shows the active parts of the five layers of pixel detectors.
The $\sim 10\,000$ hybrid modules require a fast and reliable readout at the front-end chip trigger rate of 1\,MHz, which corresponds to $\sim 25\,$Tb/s of data rate.
For the readout of the pixel modules, their electrical signal is converted to an optical signal right outside the ITk in the Optosystem.

%%%%%%%%%%%%%%%%%%%%%

\section{The Optosystem for the ITk Pixel Detector Readout}

\begin{figure}
   \centering
     \includegraphics[width=\textwidth]{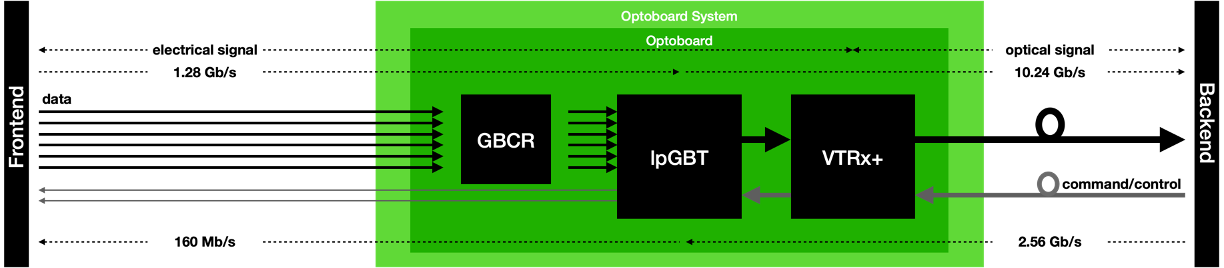}     
    \caption{\label{fig:datatransmission}Data transmission scheme with the Optoboard System \color{black}for electro-optical conversion.
    The electrical signal from the modules is transmitted with twinax copper cables and the optical signal via fibres. \color{black}
    The uplink goes through the GBCR and the lpGBT.
    On the downlink, the GBCR is bypassed.}
\end{figure}

The optical-electrical conversion \color{black}(and vice versa) \color{black} in the readout system is performed by the Optosystem, as depicted in Figure\,\ref{fig:datatransmission}.
All up- and downlinks are transmitted through this system.
The downlinks send command and trigger signals from the off-detector readout system which is built of $\sim 190$ FrontEnd LInk Exchange (FELIX)\cite{Ryu_2017,Solans:2674293} cards to the front-end chips in the pixel detector modules.
The uplinks transmit the detector signals coming from the front-end chips to the FELIX cards.
\color{black} The data transmission path starts with an electrical part where \color{black}twinax \textcolor{black}{copper} cables connect the modules to the Optoboards with a length of 6\,m at the most.
The optical fibres have a length of $\sim 150$\,m.

\begin{figure}
\centering
\begin{subfigure}[c]{0.39\textwidth}
\centering
\includegraphics[height=0.85\linewidth]{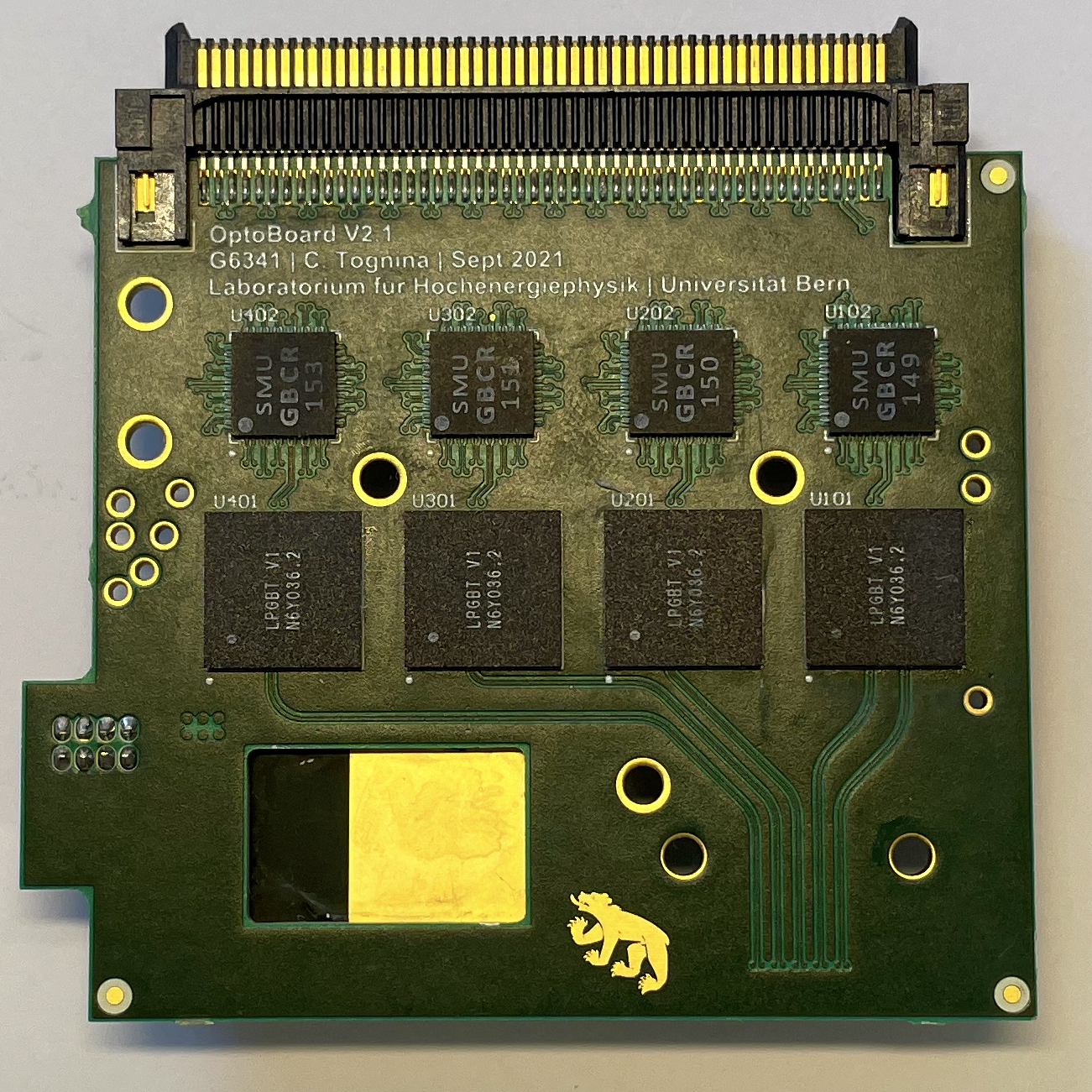}
\subcaption{\label{fig:board}Optoboard v2.1 backside with four \\lpGBTs and four GBCRs.}
\end{subfigure}
\begin{subfigure}[c]{0.6\textwidth}
\centering
\includegraphics[height=0.55\linewidth]{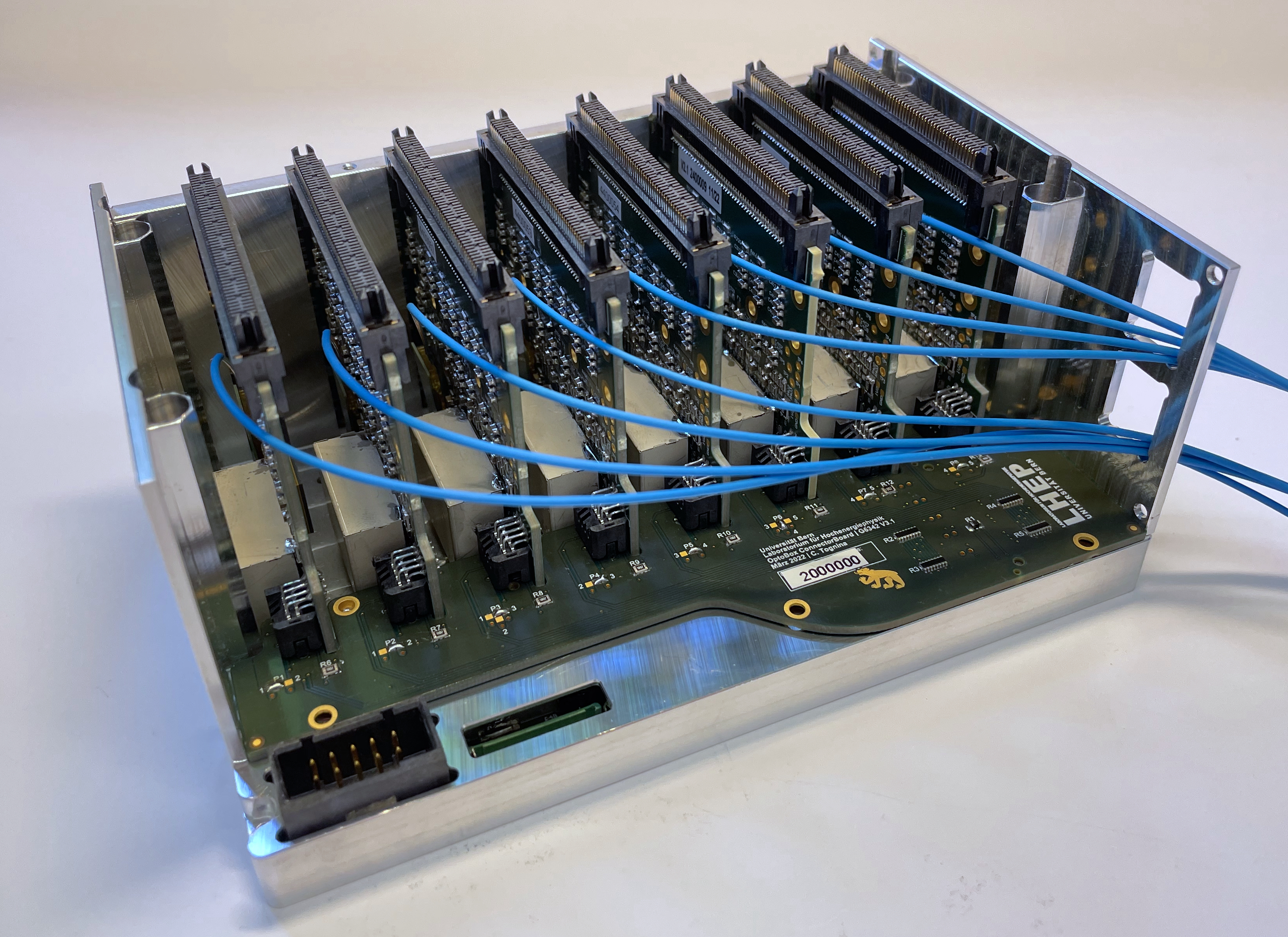}
\subcaption{\label{fig:box}Fully populated Optobox with Powerbox and connector board and open side.}
\end{subfigure}\\
\begin{subfigure}[c]{0.4\textwidth}
\centering
\includegraphics[height=0.88\textwidth]{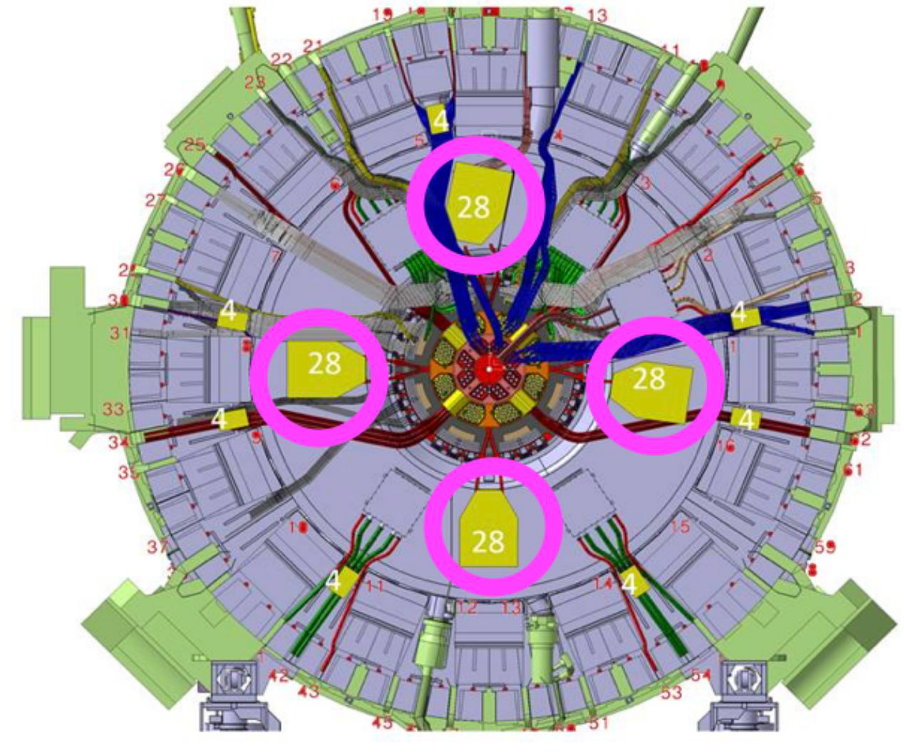}
\subcaption{\label{fig:pp1}Optopanel position in ATLAS endplate.}
\end{subfigure}
\begin{subfigure}[c]{0.59\textwidth}
\centering
\includegraphics[height=0.6\textwidth]{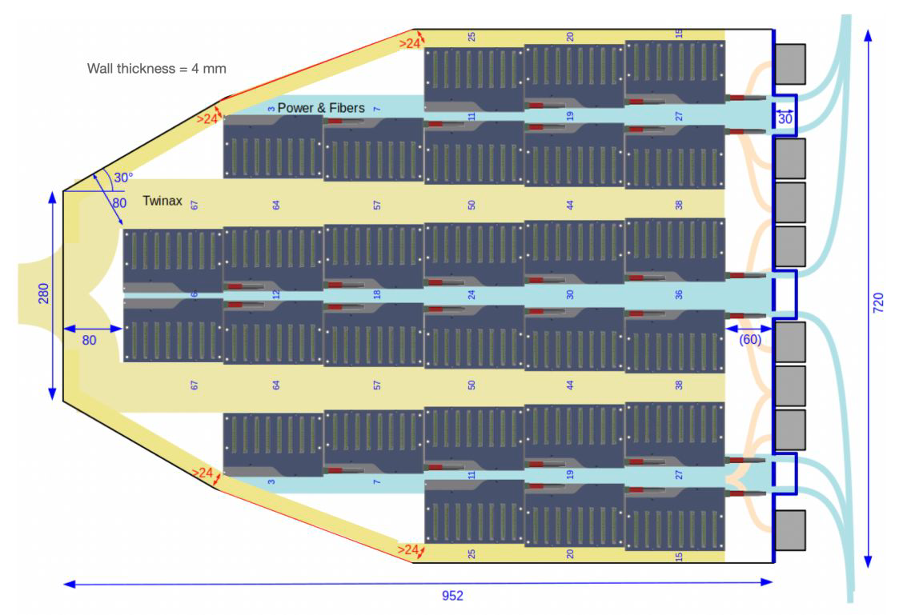}
\subcaption{\label{fig:panel}Optopanel with Optoboxes and cable channels.}
\end{subfigure}
\label{fig:optocomponents}
\caption{The position of the Optosystem in the ATLAS detector on the end plate and its components.}
\end{figure}

The centrepiece of the Optosystem is the Optoboard, a printed circuit board (PCB) with several application specific integrated circuits (ASICs) for the electrical-to-optical conversion of the signal and vice versa.
There will be $\sim 1600$ Optoboards in the ITk.
Up to 8 Optoboards are housed in a mechanical structure called Optobox, as shown in Figure\,\ref{fig:board} and \ref{fig:box}. 
The 8 Optoboards are connected via a Connectorboard to a Powerboard, which is responsible for the powering of the Optoboards and \color{black}the slow control \color{black} monitoring via a MOnitoring on Pixel System (MOPS) chip\cite{Ahmad_2022,Ahmad_2023}.
The Powerboard is shielded from electro-\color{black}magnetic \color{black} interference via an aluminium Powerbox.
The Optoboxes with Powerboxes are situated in 8 Optopanels, located on both sides of the ATLAS detector between $R=1450$\,mm and $R=2400$\,mm and at $z\approx 3500$\,mm.
They house 28 Optoboxes each, see Figure\,\ref{fig:panel}.
Cable channels run between the Optoboxes, separating the data cables (yellow) and the optical fibres, power and monitor cables (blue).

The panels provide Faraday cage shielding to protect the Optoboards from electromagnetic fields.
Apart from that they also house the cooling system for the Optoboxes which dissipate $\sim 25$\,W of heat per box.
The main heat dissipation stems from the bPOL12V\cite{bPOL12V} and the bPOL2V5\cite{bPOL2V5}, which are used for DCDC conversion to provide the required input voltages for the system, on the Powerboards and on the Optoboards, respectively.
%The bPOL2V5 maximum operating temperature is $30\degree$C.
The voltage for the Powerboards comes directly from the ATLAS service caverns at 9\,V.
It is converted to 2.5\,V on the power board with five bPOL12Vs.
The 2.5\,V is directly used on the Optoboard for the optical conversion in the VTRx+\cite{Troska:2312396,XIANG20121750} and also converted by a bPOL2V5 to 1.2\,V, which is needed by the low-power Gigabit Transceiver (lpGBT)\cite{Guettouche_2022,Moreira:2809058} and Giga Bit Cable Receiver (GBCR)\cite{gbcr, Zhang_2023} ASICs on the Optoboards.
The bPOLs are shielded to suppress any interference due to the oscillating magnetic field in its coils.

%\begin{figure}
%\centering
%\begin{subfigure}[c]{0.26\textwidth}
%\centering
%\includegraphics[height=0.7\linewidth]{Figures/optoboard-v2_1-front-vtrx-bpol}
%\subcaption{\label{fig:board}Optoboard v2.1 backside with four lpGBTs and four GBCRs.}
%\end{subfigure}
%\begin{subfigure}[c]{0.26\textwidth}
%\includegraphics[height=0.7\linewidth]{Figures/optobox-open-full-vtrx-powerbox}
%\subcaption{\label{fig:box}Fully populated Optobox with Powerbox and connector board and open side.}
%\end{subfigure}
%\begin{subfigure}[c]{0.46\textwidth}
%\centering
%\includegraphics[height=0.6\textwidth]{Figures/optopanel}
%\subcaption{\label{fig:panel}Optopanel with Optoboxes. Between the boxes run Twinax (yellow) and fibre (blue) channels.}
%\end{subfigure}
%%\begin{subfigure}[c]{0.24\textwidth}
%%\centering
%%\includegraphics[height=0.7\textwidth]{Figures/PP1}
%%\subcaption{\label{fig:pp1}Optopanel position in ATLAS endplate.}
%%\end{subfigure}
%\label{fig:optocomponents}
%\caption{The position of the Optosystem in the ATLAS detector on the end plate and its components.}
%\end{figure}

The signal recovery and equalisation of the electrical signal, coming from the pixel modules on the uplink, is done in the GBCR at 1.28\,Gb/s. 
Multiplexing (serialisation) of the independent sensor signals is done by \color{black}the \color{black} lpGBT, \color{black}where up to six uplinks, coming from the GBCR, are serialised into one 10.24\,Gb/s link. On each Optoboard there are four lpGBTs, one of which takes the master role for configuration and timing. \color{black}
\color{black}25\,\% of the bandwidth is dedicated to \color{black} Forward Error Correction (FEC) which detects and corrects \color{black}serialisation, optical conversion, transmission and reconversion errors within the lpGBT\color{black}.
On the downlink, the command and \color{black}clock reach the modules directly \color{black} from the lpGBT where they are demultiplexed into \color{black}two links. 
In total, there are eight links per Optoboard.
The downlink is transmitted at \color{black}2.56\,Gb/s.
Half of the bandwidth is used for the 8 downlinks to the FEs,  about a third for FEC and the rest for the internal slow control (IC) of the lpGBT and headers\color{black}.  
The opto-electrical conversion takes place in the VTRx+.
More than 4000 optical fibres are routed to the FELIX readout system.
The configuration of the Optoboard is using the IC protocol via the master lpGBT.
It communicates with all other slave ASICs via \color{black}the Inter-Integrated Circuit (I$^2$C) serial communication bus\color{black}. 
Data can be sent by all lpGBTs to FELIX.
These ASICs need to be radiation hard as the estimated dose of radiation during their lifetime is $\sim 50$\,kGy.

\section{Data Transmission Tests}
In order to test the data transmission capability of the Optosystem, the data transmission chain for the ITk Pixel detector is built for one single module.
\color{black}This includes a 6\,m long twinax cable, an Optoboard and a fibre connection to the FELIX readout system. \color{black}
To ensure a successful readout, it is estimated that the total attenuation of the signal \color{black}in the electrical part of the system \color{black} must be less than 20\,dB.
Also, the signal transmission is evaluated with bit error ratio tests (BERT), \color{black}where imperfect data coming from the modules is seen as bit errors\color{black}.
With these \color{black}BERT \color{black} scans different parts of the data transmission path can be verified. 
While tests with adapter cards can be used to evaluate the Optoboard performance alone, the full data transmission chain \color{black} from the module to FELIX with the final components \color{black} for any of the three subsystems can be used to estimate the overall performance.
During the development of the Optoboard, the different versions were realised.
Optoboard v0 provided a GBCR conceptual proof, v1, v2, v2.1, v3 and v4 were developed to fit the needs of the system with a fully populated board in v1, a cooling system upgrade and I$^2$C bug fixes for v2, a new lpGBT version in v2.1, a new downlink routing in v3 and halogen free PCB in v4.
Twinax cables from different vendors can be validated and compared with the full data transmission chain.
Currently the Outer Barrel subsystem components and one almost final module ITkPix v1 are used for the tests.

There are two ways of performing BERTs by configuring the module FEs differently.
The Pseudorandom Binary Sequence PRBS7 signal of the module can be used, that is in this case checked by the internal pattern checker of the lpGBT.
Another method is using the 64b/66b ITkPix idle signal, read out via optical fibre by FELIX.
The second test is on a realistic data stream and also referred to as soft error test, where 62\,\% of the frame is checked with FELIX and sensitive to errors.

\begin{figure}
\centering
\includegraphics[width=.8\textwidth,keepaspectratio]{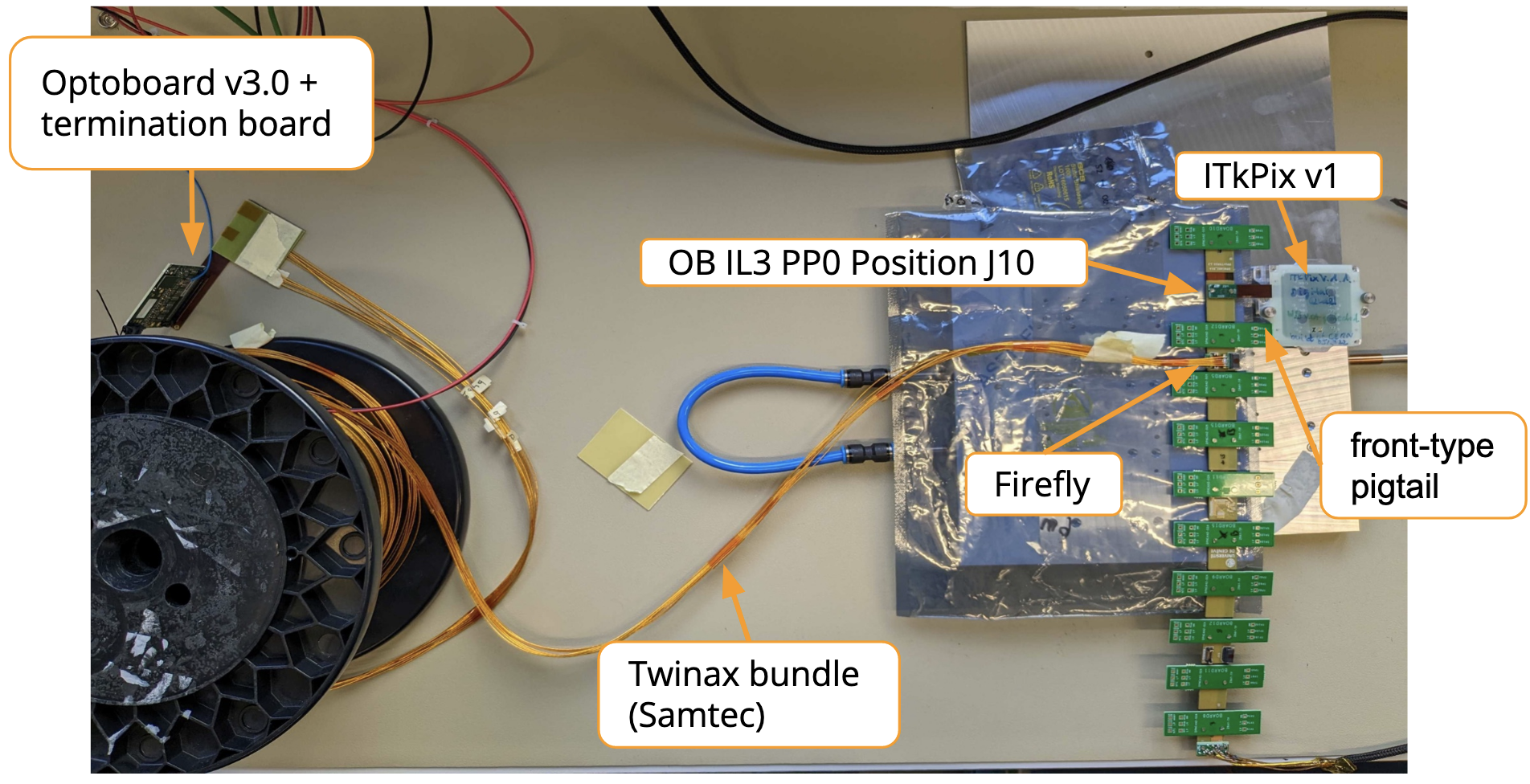}
\caption{\label{fig:QCsetup}QC setup for the data transmission tests of Optoboards at the University of Bern.}
\end{figure}

An overall bit error rate of $<2.7 \cdot 10^{-12}$ with a 95\% confidence level is the limit that can be set by using the pattern checker of the lpGBT.
The number of bits that can be checked with the lpGBT is insufficient to reach a lower limit. 
In the case of the soft error scans, prolongation of the monitoring time makes testing to $10^{-12}$ feasible, which is a well-established reference value used in industry. 
The setup for the latest data transmission chain tests is shown in Figure\,\ref{fig:QCsetup}.

\begin{figure}
\centering
\includegraphics[width=.49\textwidth,keepaspectratio]{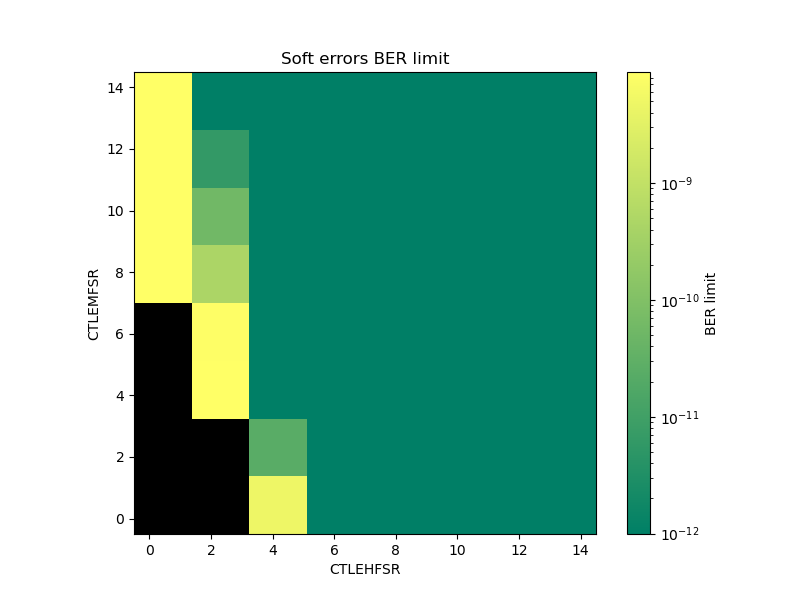}\hfill
\includegraphics[width=.49\textwidth,keepaspectratio]{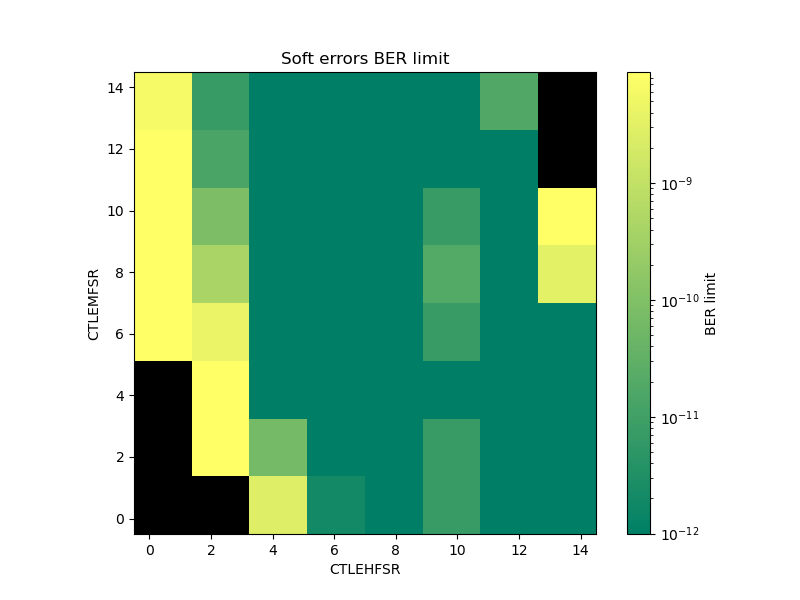}
\caption{\label{fig:Softerror}Soft error tests with two twinax bundles \color{black}from two different vendors. Vendor 1 \color{black} bundle on the left and \color{black}vendor 2 \color{black} bundle on the right.
The dark green areas show register settings with good data transmission properties.
The black areas show where no decoding alignment is possible.}
\end{figure}

Multiple soft error tests are performed with an Optoboard v3, by changing the parameters of the equaliser of the GBCR. 
For the plots shown in Figure\,\ref{fig:Softerror}, each point has been monitored for 3800\,s which corresponds to \color{black}3800\,s\,$\times \,1.28$\,Gb/s\,$\times \,62$\% \color{black} monitored bits, which are enough to set the BER limit with 95\% CL to $10^{-12}$ if no errors occur. 
The two registers of the GBCR are responsible for the equalisation of the middle (200-400\,MHz) and high frequencies (400-2000\,MHz).
According to the results there are several register combinations yielding the required BER performance, as shown in Figure\,\ref{fig:Softerror} in the dark green areas.
The black areas show register settings that lead to the signal being so distorted, that FELIX cannot establish decoding alignment as the header is not recognised.

\section{Radiation Hardness Tests \color{black}and System Tests\color{black}}
To assure the radiation hardness of the final components of the Optosystem, further irradiation tests with the Bern cyclotron (18\,MeV protons) will be conducted.
The components of the data transmission chain \color{black}will be \color{black} irradiated and their performance subsequently evaluated.
Another test is the operation of an Optoboard during irradiation.
Results of irradiation tests on previous versions of the Optosystem have been presented in \cite{Halser_2022,Santo_2023}.

In order to test system aspects, 11 sites have been appointed to build and test sections of the three subsystems of the ITk.
Later, sites from all subsystems will integrate their test setups in one large setup. 
\color{black}This final step is referred to as integration test. \color{black}
To facilitate the testing, user-friendly software for the configuration of Optoboards will be used, which comes in docker containers.
Also, a custom Optopanel with 1-4 Optoboxes for conceptual tests will be provided to the sites.

Preceding these tests, the powering scheme concept with the bPOLs was verified and the Optopanel cooling successfully tested with heat pads, mimicking the power consumption of the Optoboxes.

%%%%%%%%%%%%%%%%%%%%%%%%%%%%%%%%%%%%%%%%%
\section{Outlook and Conclusions}
The Optosystem has been designed in a way that it can handle the challenging conditions in the HL-LHC.
With the final data transmission chain the recently conducted tests confirmed that it is working within required specifications.
To further consolidate the Optosystem performance, extensive tests of the final readout path for all three subsystem will be conducted during ITk detector integration.

%%%%%%%%%%%%%%%%%%%%%%%%%%%%%%%%%%%

%\appendix
%\section{Some title}
%Please always give a title also for appendices.

%\acknowledgments
%
%This is the most common positions for acknowledgments. A macro is
%available to maintain the same layout and spelling of the heading.

%\paragraph{}

% We suggest to always provide author, title and journal data:
% in short all the informations that clearly identify a document.

%\begin{thebibliography}{99}

%\bibitem{a}
%ATLAS Collaboration, \emph{The ATLAS Experiment at the CERN Large Hadron Collider}, \emph{J. Abbrev.} {\bf vol} (year) pg.
%
%\bibitem{b}
%Author, \emph{Title},
%arxiv:1234.5678.
%
%\bibitem{c}
%Author, \emph{Title},
%Publisher (year).
\bibliographystyle{JHEP}
\bibliography{lit}

% Please avoid comments such as "For a review'', "For some examples",
% "and references therein" or move them in the text. In general,
% please leave only references in the bibliography and move all
% accessory text in footnotes.

% Also, please have only one work for each \bibitem.

%\end{thebibliography}
\end{document}